# Knowledge Elecitation for Factors Affecting Taskforce Productivity– using a Questionnaire

Muhammad Sohail*
Institute of Information Technology
Kohat University of Science & Technology (KUST)
Kohat, Pakistan
sohailshinwari76@yahoo.com

Abdur Rashid Khan
Institute of Computing & Information Technology
Gomal University
Dera Ismail Khan, Pakistan
rashidkh08@yahoo.com

*Abstract*—in this paper we present the process of Knowledge Elicitation through a structured questionnaire technique. This is an effort to depict a problem domain as "Investigation of factors affecting taskforce productivity". The problem has to be solved using the expert system technology. This problem is the very first step how to acquire knowledge from the domain experts. Knowledge Elicitation is one of the difficult tasks in knowledge base formation which is a key component of expert system. The questionnaire was distributed among 105 different domain experts of Public and Private Organizations (i.e. Education Institutions, Industries and Research etc) in Pakistan. A total of 61 responses were received from these experts. All the experts were well qualified, highly experienced and has been remained the members for selection committees a number of times for different posts. Facts acquired were analyzed from which knowledge was extracted and elicited. A standard shape was given to the questionnaire for further research as a knowledge acquisition tool. This tool may be used as a standard document for selection and promotion of employees.

*Keywords- Expert System; Knowledge Acquisition; Knowledge Elicitation; Questionnaire; Taskforce Productivity; domain experts*

## I. INTRODUCTION

In mid of sixties Artificial Intelligent community developed the Expert Systems technology [16]. Expert systems were consists of tools used for decision making using the reasoning mechanisms of human expert in their area of expertise [1]. The use and development of expert system is very much in large number in different field of life. The knowledge base is the important component of the expert system [9]. The knowledge acquisition (KA) is the process of gathering and transformation of problem solving expertise from human expert through computer program [16]. The process of KA from Expert of the domain some how also called Knowledge Elicitation. The Knowledge Acquisition and Elicitation is the one of the difficult task of the expert system development [1, 11, 12, 16, and 17].

There are a number of KA techniques we found from literature [1, 8, 11, 12, and 16]. But the selection of KA technique depends upon the nature of problem, while accessibility and availability of the domain expert is another aspect to be considered. In the problem domain questionnaire was used as a knowledge acquisition tool. This technique is the most suitable if knowledge has to be acquired from domain experts when reasonable period can be spared.

To solve the problem of "Finding factors affecting taskforce productivity", domain experts were selected from Universities, Industries, Semi-Government and Private Organizations in Pakistan.

## II. KNOWLEDGE ACQUISITION & ELICITATION

Knowledge elicitation is known to be one of the major bottle-necks in the knowledge base development [11].

Acquiring knowledge from human experts and learning from data is the knowledge elicitation [6]. The knowledge engineer works with the domain expert having the expertise in the specific domain area. The knowledge engineer applies the knowledge elicitation techniques to acquire knowledge. After that the knowledge is to be coded in computer format further to be used as a knowledge base for inferences, decisions and getting new knowledge.

### A. Knowledge Elicitation Techniques

The knowledge Elicitation techniques are used by the knowledge engineer to acquire knowledge from human experts to solve problem. There are a number of Knowledge elicitation techniques, for example the interview, structured interview, questionnaire, protocol analysis, concept sorting, simulation and prototyping [1, 6, 7, and 17]. These techniques are adopted according to the nature of the problem.

## III. APPROACH

The objective of this study is to find out those factors which truly affect the productivity of the taskforce. This needed a complete knowledge elicitation process. This study was completed through the steps like; *Problem Identification, Domain Expert Selection, Adaptation of suitable Knowledge Elicitation Technique, Questionnaire design and distribution, Analysis* and finally the *Conclusion*. In the problem domain experts were available only at remote areas of the country. Due to the basic nature of the problem domain a questionnaire was selected for knowledge acquisition and elicitation.

---

* This is a part of my MS research work

### A. Problem Identification

The first step in knowledge elicitation was to identify the factors affecting taskforce productivity. Various Organizations' Annual Confidential Reports (ACR) was studied, domain experts were contacted and factors were chose for a complete crude questionnaire.

Literature study reveals that the productivity is directly proportional to the selection process. In order to control the human resources and quality in human resource management, the right talent selection is important. [14, 15]

With an effective personal selection process the organization can improve their productivity problem. It can be happen by selecting right talent for the right job.

Ref [2] stated that the factors like employee's age, gender, marital status, educational background, and work experience predict the employee work performance and retention. They also emphasizes on the proper orientation and management of the selection process carried out by the company's human resource department. Motivation theories emphasizes on the employee's basic needs such as food, drink, sleeping hours, and on broader scenario the factors like security, love, status, self respect, growth and accomplishment etc. Factors like the supervision, relationship with the supervisors, salary, working conditions, company policy and administration, all affecting the productivity of organization and the individual performance[3,4,7].

Thus there is a need of a systematic and strong intelligent knowledge base that must assist us in selecting the right talent for right job and further to be used for decision making for the improvement of individual work performance.

### B. The Selection Process

Ref [2] has reviewed that factors like person, organization, society, law, market influence, the nature and analysis of work behavior, and information technology achievements, affecting the selection process and the productivity of the human resources.

The application of expert system or Decision Support System on selection process for right talent chasing is increasing [5].

The organization objective are very simple, the right talent selection for long period of time giving improvements to the current status. Therefore the right talent selection is a goal to apply a valid and effective method to reduce the risk in selecting unsuitable person and increasing the opportunities to find an eligible employee who can enhance the productivity of the organization [2, 3, 7].

### C. The Domain Expert Selection

The next task was to select the experts of the problem area for knowledge elicitation and problem solution. The collection of expert knowledge for the knowledge base is the main task of the expert system development, and the important role is of the human expert in this scenario [9].

The domain problem is related with the areas of Psychology, Human Resource Management, Education, Industries, and almost all professions.

In order to set standard factors affecting taskforce productivity these experts were traced through **140** organizations of Pakistan. We found **112** domain experts along with their relevant information, experience, qualifications, contact addresses and areas of interests. Here one important issue was to find domain experts. They were found both in service and/or recently retired from their services. It is because the domain expert knowledge is the main objective [9]. See Table I for details.

TABLE I. CHARACTERISTICS OF DOMAIN EXPERTS

| Domain Experts |
|---|
| **Gender** |
| – Male |
| – Female |
| **Designation & Nature of Job** |
| – Doctors |
| – Engineers |
| – Managers |
| – Administrators |
| – Academisians |
| – Professors |
| – Judges |
| **Qualification** |
| – Post Doc |
| – PhD |
| – M.Phil |
| – Master |
| – Bachellors (MBBS, Engineers) |
| – Others |
| **Subject Specialty** |
| – Law |
| – Engineering |
| – Education |
| – Human Resource Management |
| – Psychology |
| – Computer Science |
| – Statistics |
| – Health |
| – Agriculture |
| – Others |
| **Demographics** |
| – Federal |
| – Sindh |
| – Punjab |
| – NWFP |
| – Balochistan |

Therefore the experts of those areas are traced and contacted through various means of communication, like through personnel contacts, ordinary posts and emails.

### D. Selection of Knowledge Elicitation Technique

To acquire knowledge the knowledge elicitation technique has to be applied. The literature study revealed that a number of techniques exist for knowledge elicitation, like structured interview, questionnaire, protocol analysis, concept sorting, simulation and prototyping [1, 6, 7, and 17].

*E. Questionnaire Design and Distribution*

The main purpose of the questionnaire was to sort out those factors/competencies that affect the taskforce productivity. Blank Annual Confidential Report (ACRs), progress evaluation techniques, selection and promotion criteria's were give due consideration during questionnaire development. Experts' opinions of various fields of studies, like, Human Resource Management, Psychology, Education and Computer Science were taken into account. At last the questionnaire shaped into a more than 100 competencies / factors with five fuzzy logic variables and scales, defined as: 5= Strongly Agree, 4= Agree, 3= Neutral, 2= Disagree and 1= Strongly Disagree (APPENDIX-A). First questionnaire was checked by the linguistic experts. Some factors were found to be of the same meanings and related grades, therefore changes were made accordingly and resultantly achieved **67** factors. Questionnaire was distributed among **105** experts for evaluation to finalize the importance of the decision making factors. These experts were given rights of the insertion, deletion and editing of factors along with the grading to the fuzzy variables. A time of more than three months were given along with reminder from time to time. We got **61** experts opinions after four months struggle. Study resulted that **57** competencies and factors were of importance. Therefore we evaluated them and set them for further analysis. Original data may be seen in the MS thesis research report.

*F. Analysis and findings*

We got **61** experts opinions in the form of filled questionnaire with editing and suggestions. Final questionnaire was achieved after analysis as most of the factors were found very important while some of them were relatively of lesser importance. For example the factor, interest in job was found critically sound and of very importance than the reasoning capability factor. See TABLE II (Summary of the factors affecting taskforce productivity as APPENDIX-B). Similarly APPENDIX-C graphically depicts the analysis of experts' opinions of the first and last five competencies.

## IV. CONCLUSION AND RECOMMENDATIONS

The resultant questionnaire may be used as a standard model for promotion, job redesign, job rotation, monitoring & control and selection of employees, leading to smooth operations and sustained growth of governmental and private organizations.

It was concluded that if a questionnaire is properly designed and a systematic process is adopted then knowledge from multiple experts can be extracted easily even living at remote places. This tool may be used as a pedagogical device for students and researchers. This work is not limited to Pakistan but can be extended to all countries in the world.

AUTHOR'S PROFILE

**Muhammad Sohail**

The author is currently pursuing his MS degree in Computer Science from the Institute of Information Technology, Kohat University of Science & Technology (KUST), Kohat, Pakistan. His area of interest includes; Expert System, DSS, Databases and Data Mining.

**Abdur Rashid Khan**

The author is presently working as an Associate Professor at ICIT, Gomal University D.I.Khan, Pakistan. He received his PhD degree from Kyrgyz Republic in 2004. His research interest includes AI, Software Engineering, MIS, DSS and Data Bases.

**APPENDIX-A**

A SURVEY TO INVESTIGATE FACTORS AFFECTING TASK FORCE PRODUCTIVITY

The purpose of this study is to explore the personal competency that affects job performance. Your feedback would be the base of our research. Therefore, it would be highly appreciated if you could kindly provide us the accurate / right information well in time to enable us to contribute our research work. Moreover, the information provided by you would be kept confidential and in completely secret.

Name: _____________________ Designation: _____________________

Address: _____________________

Qualification: _____________________ Gender: _____________________

Age: _________ Field: _____________ Experience: _____________

| No | FACTORS | Strongly Disagree (1) | Disagree (2) | Neutral (3) | Agree (4) | Strongly Agree (5) |
|----|---------|-----------------------|--------------|-------------|-----------|--------------------|
| 1  | Power of thinking & logic | | | | | |
| 2  | Analytical skill | | | | | |
| 3  | Power of creative thinking | | | | | |
| 4  | Ability to learn and understand | | | | | |
| 5  | Self confidence | | | | | |
| 6  | Ability to work in groups | | | | | |
| 7  | Ability to work independently | | | | | |
| 8  | Labor commitment | | | | | |
| 9  | Decision-making ability under different situations | | | | | |
| 10 | Academic record | | | | | |
| 11 | Highly qualified | | | | | |
| 12 | Presentation and communication skills | | | | | |
| 13 | Regular and punctuality | | | | | |
| 14 | Cooperation | | | | | |
| 15 | Ability to solve problems | | | | | |
| 16 | Ability to motivate others for work | | | | | |
| 17 | Makes sacrifices for others | | | | | |
| 18 | Patriotism and love for country and humanity | | | | | |
| 19 | Care for the rights of others | | | | | |
| 20 | Leadership qualities | | | | | |
| 21 | Judgments and problem understanding | | | | | |
| 22 | Abide by rules and regulations | | | | | |
| 23 | Power of control distribution among others | | | | | |
| 24 | Behaviors as a whole | | | | | |
| 25 | Family background | | | | | |
| 26 | Job knowledge | | | | | |
| 27 | Awards | | | | | |
| 28 | Grants | | | | | |
| 29 | Resources Utilization | | | | | |
| 30 | Power of honesty | | | | | |
| 31 | Power of sincerity | | | | | |
| 32 | Working environment (hotness, cold, peaceful, safety) | | | | | |
| 33 | Availability of basic needs and incentives | | | | | |
| 34 | Working efficiency and effectiveness | | | | | |
| 35 | Care for standards | | | | | |
| 36 | Professional approach to work | | | | | |
| 37 | Keep his promise | | | | | |
| 38 | Secrete Trust | | | | | |
| 39 | Interest in job | | | | | |
| 40 | Job satisfaction | | | | | |
| 41 | Ability to utilize opportunity of job trainings and courses etc | | | | | |
| 42 | Strong health physique | | | | | |
| 43 | Want to live in a simple way or want luxuries | | | | | |
| 44 | Working ability (hardworking or not?) | | | | | |
| 45 | Religious minded | | | | | |
| 46 | Liberal minded | | | | | |

*(IJCSIS) International Journal of Computer Science and Information Security,*
*Vol. 3, No. 1, 2009*| No | FACTORS | Strongly Disagree (1) | Disagree (2) | Neutral (3) | Agree (4) | Strongly Agree (5) |
|---|---|---|---|---|---|---|
| 47 | Careful | | | | | |
| 48 | Work experience | | | | | |
| 49 | Knowledge of various technologies | | | | | |
| 50 | Power of tolerance and patience | | | | | |
| 51 | Power of awareness | | | | | |
| 52 | Technical Skills | | | | | |
| 53 | Information sharing capability | | | | | |
| 54 | Management skill | | | | | |
| 55 | Planning skills | | | | | |
| 56 | Stress management | | | | | |
| 57 | Abide by rules and regulations | | | | | |

**APPENDIX-B**

TABLE II. RESPONSE SUMMARY FOR FACTORS AFFECTING TASKFORCE PRODUCTIVITY

(Total responses N= 61, 5= Strongly Agree, 1= Strongly Disagree)

| Factors | Mean | Std. Deviation |
|---|---|---|
| Interest in the job | 4.74 | 0.4435 |
| Power of sincerity | 4.66 | 0.4791 |
| Secrete Trust | 4.66 | 0.4791 |
| Professional approach to work | 4.66 | 0.4791 |
| Care for standards | 4.64 | 0.4842 |
| Keep his promise | 4.64 | 0.4842 |
| Knowledge of various technologies | 4.62 | 0.5217 |
| Power of thinking, logic | 4.61 | 0.4926 |
| Availability of basic needs and incentives | 4.61 | 0.6653 |
| Working ability (hardworking or not?) | 4.57 | 0.4986 |
| Planning skills | 4.57 | 0.5310 |
| Abide by rules and regulations | 4.56 | 0.5331 |
| Job satisfaction | 4.56 | 0.5635 |
| Leadership qualities | 4.56 | 0.5635 |
| Technical Skills | 4.56 | 0.8067 |
| Ability to utilize opportunity of job trainings and courses etc | 4.54 | 0.7433 |
| Information sharing capability | 4.51 | 0.5361 |
| Job knowledge | 4.51 | 0.5951 |
| Stress management | 4.51 | 0.6487 |
| Power of awareness | 4.49 | 0.5951 |
| work experience | 4.49 | 0.6982 |
| Judgment and problem understanding | 4.49 | 0.7664 |
| Makes sacrifices for others | 4.49 | 0.7879 |
| Resources Utilization | 4.48 | 0.6978 |
| Working efficiency and effectiveness | 4.48 | 0.8681 |
| Power of creative thinking | 4.46 | 0.8078 |
| Ability to solve problems | 4.44 | 0.8471 |
| Awards Holder | 4.41 | 0.9377 |
| Working environment (hotness, cold, peaceful, safety etc) | 4.39 | 0.8996 |
| Decision-making ability under different situations | 4.38 | 0.7340 |
| Patriotism and love for country and humanity | 4.36 | 0.8570 |
| Self confidence | 4.36 | 0.8950 |
| Academic record | 4.36 | 0.9135 |
| Cooperation | 4.34 | 0.8344 |
| Family background | 4.34 | 0.8344 |
| Power of tolerance and patience | 4.34 | 0.9108 |
| Power of honesty | 4.33 | 0.8108 |
| Careful | 4.33 | 1.1651 |
| Management skill | 4.31 | 0.8858 |
| Grant producer/Receiver | 4.28 | 1.0022 |
| Highly qualified | 4.26 | 0.9815 |
| Labor commitment | 4.26 | 1.0149 |
| Care for the rights of others | 4.26 | 1.0472 |
| Regular and punctuality | 4.25 | 1.0433 |
| Ability to motivate others for work | 4.23 | 1.0707 |
| Presentation and communication skills | 4.21 | 1.0347 |
| Ability to work independently | 4.15 | 0.8131 |
| Behavior as a whole | 4.13 | 1.1471 |
| Analytical skill | 4.08 | 1.3576 |
| Ability to learn and understand | 4.07 | 0.9810 |
| Religious minded | 4.05 | 1.2169 |
| Ability to work in groups | 4.03 | 1.1250 |
| Liberal minded | 3.92 | 1.1874 |
| Power of control distribution among others | 3.82 | 1.2180 |
| Strong health physique | 3.49 | 1.4677 |
| Want to live in a simple way or want luxuries | 3.03 | 1.6730 |
| Reasoning Capability | 3.03 | 1.6730 |

**APPENDIX-C**
Graphic Analysis of Experts' Opinions

**Expert's Opinion in Percentage**

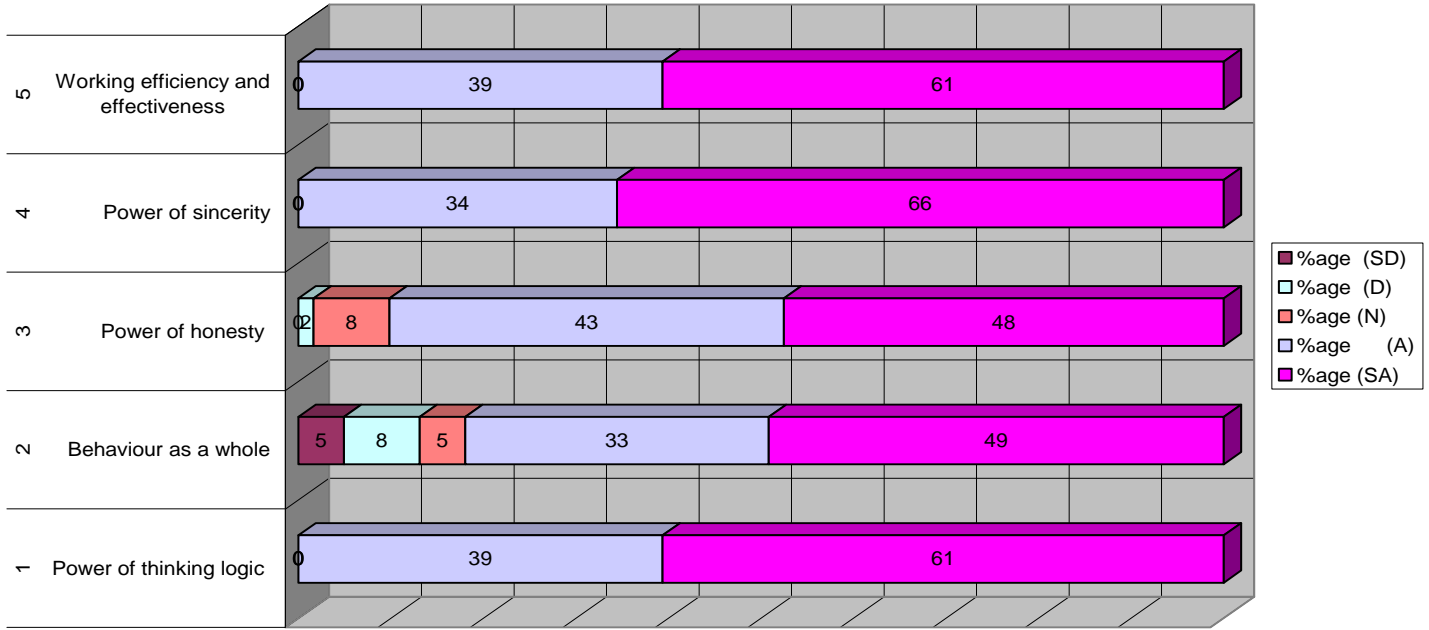

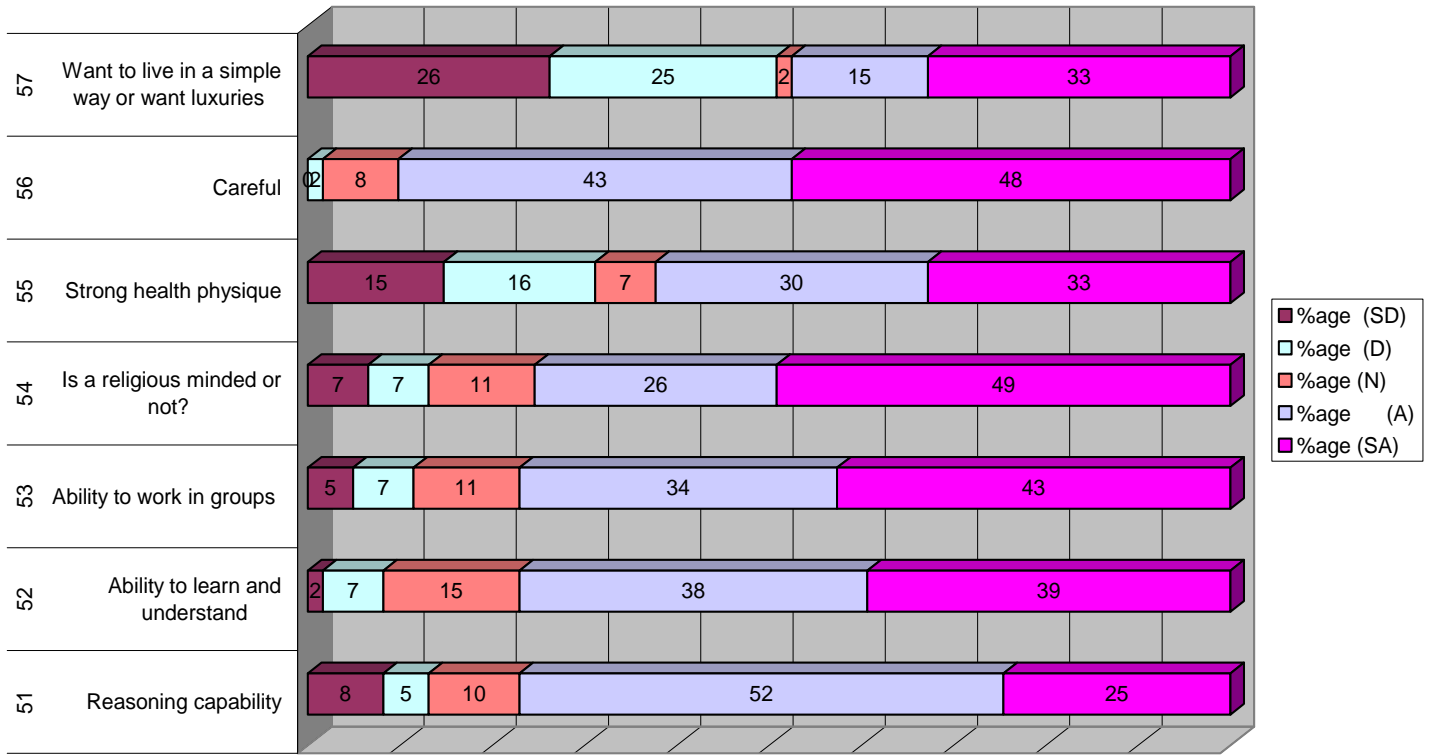